\documentclass[twocolumn]{aastex702}

\usepackage{amsmath}
\usepackage{graphicx}

\shorttitle{Dynamical microlensing and disk-size bias}
\shortauthors{Fagin et al.}

\begin{document}

\title{Neglecting Dynamical Microlensing Can Bias Accretion Disk Size Measurements\\
at the Population Level}

\author[0000-0001-8723-6136]{Joshua~Fagin}
\affiliation{Department of Physics and Astronomy, Stony Brook University, Stony Brook, NY 11794, USA}
\email[show]{joshua.fagin@stonybrook.edu}
\affiliation{The Graduate Center of the City University of New York, 365 Fifth Avenue, New York, NY 10016, USA}
\affiliation{Department of Astrophysics, American Museum of Natural History, Central Park West and 79th Street, NY 10024-5192, USA}
\affiliation{Department of Physics and Astronomy, Lehman College of the CUNY, Bronx, NY 10468, USA}

\author[0000-0001-8797-725X]{James~Hung-Hsu~Chan}
\affiliation{Department of Astrophysics, American Museum of Natural History, Central Park West and 79th Street, NY 10024-5192, USA}
\affiliation{Department of Physics and Astronomy, Lehman College of the CUNY, Bronx, NY 10468, USA}
\email[]{}

\author[0000-0002-9086-6398]{Connor~Stone}
\affiliation{Dunlap Institute, University of Toronto, Toronto, Canada}
\affiliation{Canadian Data Intensive Astrophysics PLatform (CanDIAPL)}
\affiliation{Ciela - Montreal Institute for Astrophysical Data Analysis and Machine Learning, Montr\'eal, Canada}
\affiliation{Department of Physics, Université de Montréal 2900 Edouard Montpetit Blvd Montréal Québec H3T 1J4, Canada}
\affiliation{Mila - Quebec Artificial Intelligence Institute, Montr\'eal, Canada}
\email[]{}

\author[0009-0009-6932-6379]{Henry~Best}
\affiliation{Department of Physics and Astronomy, Stony Brook University, Stony Brook, NY 11794, USA}
\email[]{}

\author[0000-0002-6909-205X]{Paras~Sharma}
\affiliation{Department of Physics and Astronomy, Stony Brook University, Stony Brook, NY 11794, USA}
\email[]{}

\author[0000-0003-0631-9701]{Sophia~Miskiewicz}
\affiliation{The Graduate Center of the City University of New York, 365 Fifth Avenue, New York, NY 10016, USA}
\affiliation{Department of Astrophysics, American Museum of Natural History, Central Park West and 79th Street, NY 10024-5192, USA}
\affiliation{Department of Physics and Astronomy, Lehman College of the CUNY, Bronx, NY 10468, USA}
\email[]{}

\author[0009-0000-4476-5003]{Matthew~O'Dowd}
\affiliation{The Graduate Center of the City University of New York, 365 Fifth Avenue, New York, NY 10016, USA}
\affiliation{Department of Astrophysics, American Museum of Natural History, Central Park West and 79th Street, NY 10024-5192, USA}
\affiliation{Department of Physics and Astronomy, Lehman College of the CUNY, Bronx, NY 10468, USA}
\email[]{}

\author[0000-0003-3544-3939]{Laurence~Perreault-Levasseur}
\affiliation{Ciela - Montreal Institute for Astrophysical Data Analysis and Machine Learning, Montr\'eal, Canada}
\affiliation{Department of Physics, Université de Montréal 2900 Edouard Montpetit Blvd Montréal Québec H3T 1J4, Canada}
\affiliation{Mila - Quebec Artificial Intelligence Institute, Montr\'eal, Canada}
\affiliation{Trottier Space Institute, Montr\'eal, Canada}
\email[]{}

\author[0000-0002-8669-5733]{Yashar~Hezaveh}
\affiliation{Ciela - Montreal Institute for Astrophysical Data Analysis and Machine Learning, Montr\'eal, Canada}
\affiliation{Department of Physics, Université de Montréal 2900 Edouard Montpetit Blvd Montréal Québec H3T 1J4, Canada}
\affiliation{Mila - Quebec Artificial Intelligence Institute, Montr\'eal, Canada}
\affiliation{Trottier Space Institute, Montr\'eal, Canada}
\email[]{}

\author[0000-0003-3195-5507]{Simon~Birrer}
\affiliation{Department of Physics and Astronomy, Stony Brook University, Stony Brook, NY 11794, USA}
\email[]{}

\begin{abstract}
Microlensing of strongly lensed quasars probes unresolved accretion disks, but many inference methods use static magnification maps with fixed relative microlens positions. Using the GPU-accelerated simulator \textsc{Microcaustics}, we generate ten-year multiband light curves for three quadruply lensed quasar configurations. We compare dynamical microlensing with independent microlens motions against coherent bulk motion, and a microlens mass distribution against equal masses, keeping the velocity distribution of individual microlenses and mean mass fixed. We infer population disk sizes using ridge regression on the means and lower and upper scatters of ten variability and event metrics. We measure biases by applying models trained with simplified physics to simulations including both effects. Models trained on coherent motion interpret the additional variability from evolving caustics as smaller disks. For populations of 128 independent realizations, jointly inferred radii are consistently smaller by 6--28\%. Microlens dynamics increase caustic-crossing rates pooled across all four macroimages by 5--15\% and $r$-band high-magnification-event (HME) rates by 13--25\%, while shortening median HME durations by 4--12\%. Omitting the mass distribution produces configuration-dependent disk-size offsets from $-12\%$ to $+5\%$. Including the mass distribution decreases caustic-crossing rates by 6--8\% and HME rates by 6--12\%. It also narrows the spread of ten-year averaged microlensing time-delay differences between macroimages by a median of 9--12\% per system, while microlens dynamics have a smaller, image-pair-dependent effect. These results demonstrate the importance of microlens dynamics for population disk-size inference and motivate simulations that include dynamical microlensing and a microlens mass distribution.
\end{abstract}

\keywords{\uat{Quasar microlensing}{1318} --- \uat{Accretion}{14} --- \uat{Quasars}{1319} --- \uat{Strong gravitational lensing}{1643}}

\section{Introduction}
\label{sec:introduction}

Gravitational microlensing of strongly lensed quasars provides a way to measure accretion disks on scales far below the angular resolution of current telescopes \citep{AvilaVera2026}. Stars and other compact objects in a foreground lens galaxy produce a network of caustics that differentially magnifies the unresolved background source. A larger emitting region averages over more of this structure, so the amplitudes and timescales of microlensing fluctuations carry information about its spatial extent. Because the disk size changes with wavelength, multiband microlensing observations also probe its temperature profile \citep{Poindexter2008,Blackburne2011}. Caustic crossings can additionally probe the relativistic structure of the inner disk \citep{Mediavilla2015InnerDisk,Best_2024}. Microlensing measurements have often favored disks larger than expected from their luminosities under the standard thin-disk model, motivating tests of both accretion-disk physics and the assumptions entering microlensing inference \citep{Shakura_1973,Morgan2010,Blackburne2011,Cornachione_2020}.

A common approximation is to move the source through a static magnification map, such as the GERLUMPH maps~\citep{Vernardos2014}. This captures relative bulk motion between the observer, lens, and source \citep{Kayser1986}, but holds the positions of the microlenses fixed relative to one another. In a real lens galaxy, stellar proper motions continuously change their separations and reshape the caustics. Early moving-star calculations showed that this evolution can produce events that are shorter, steeper, and more frequent than those generated by bulk motion alone \citep{KundicWambsganss1993,KundicWittChang1993,WambsganssKundic1995}. An effective transverse velocity can reproduce some statistical consequences of microlens motions~\citep{Wyithe2000Derivatives}.

While the majority of microlensing analyses have used static microlensing maps due to practical computational limitations, microlens motions have already been included in microlensing inference. \citet{Poindexter10Methods,Poindexter10} used evolving magnification patterns in Bayesian analyses of Q2237+0305 to constrain the lens velocity, mean stellar mass, and accretion-disk structure. Evolving magnification patterns have also been used to infer quasar emission-region sizes and structure in subsequent multiband analyses \citep{Mosquera2013,Blackburne2014,Blackburne2015}. Source size, microlens mass, and velocity all influence the observed variability. A model that changes the temporal structure of a light curve can consequently favor a different source size even when the underlying disk is unchanged. How this changes the inferred disk size depends on the statistics and inference method used.

A second common approximation is to assign every microlens the same mass. A single mass defines one Einstein-radius scale, proportional to the square root of that mass, whereas a mass distribution supplies a range of lensing scales. Using a single mass simplifies inference by avoiding assumptions or parameters describing the mass distribution. However, magnification statistics can depend on the microlens mass distribution \citep{Schechter2004MassSpectrum,Congdon2007}, and quasar microlensing has also been used to constrain the masses and abundances of stars and other compact objects in lens galaxies~\citep{JimenezVicenteMediavilla2019,Mediavilla2017PBH,EstebanGutierrez2020,Tuntsov2024Planets}. Holding the mean mass fixed provides a useful controlled comparison, but does not guarantee that replacing a mass distribution by one mass preserves the light-curve statistics \citep{EstebanGutierrez2020}. We therefore examine how microlens motion and the microlens mass distribution each affect microlensing observables.

Continuum reverberation probes disk structure through the delayed response of disk emission to changes in the irradiating source \citep{Cackett2007}. Different parts of an irradiated disk respond at different light-travel delays \citep{chan2024reverberationmappinglamppostwind}, so magnifying these regions differently changes the mean response lag of each image \citep{best2024amoebaagnmodeloptical,TieKochanek2018,Chan2021MicrolensingDelays}. Microlens motions can make these shifts evolve over time even when the disk's mean structure remains fixed. These image-dependent response delays can affect time-delay cosmography \citep{Chen2018MicrolensingDelays}, which uses the relative arrival times of lensed images to constrain cosmological distances and $H_0$ \citep{Refsdal1964,Birrer2024TimeDelayCosmography}. The decade-long monitoring of the Vera C. Rubin Observatory's Legacy Survey of Space and Time (LSST) motivates examining how these lag distributions, as well as the microlensing variability, change with monitoring duration \citep{Ivezic2019LSST,fagin2024predictinghighmagnificationevents}. LSST is expected to discover thousands of lensed quasars \citep{OguriMarshall2010,Abe2025Lenses}, with dozens of high-magnification events (HMEs) expected annually across the monitored lens population \citep{Neira2025HME}.

Machine-learning approaches already use quasar light curves to infer black hole and disk properties \citep{VernardosTsagkatakis2019,Best_2024,Fagin_2024,Fagin2025Reprocessing}, forecast microlensing HMEs \citep{fagin2024predictinghighmagnificationevents}, and identify caustic crossings and infer source parameters (J. H.-H. Chan et al. 2026, submitted). Disk sizes have also been inferred from statistical light-curve features using supervised regression \citep{JimenezVicenteMediavilla2025ML}. These methods depend on the physics included in their training simulations, so omitted microlens motions or a simplified mass distribution could bias their predictions.

We test whether neglecting relative microlens motion or replacing the microlens mass distribution by a single mass biases population-level disk-size inference. We use \textsc{Microcaustics} \citep{Fagin2026Microcaustics}, which enables efficient large-scale dynamical microlensing simulations with a mass distribution and physically motivated general-relativistic accretion disk models. Four simulation families vary these two assumptions independently, using the same source models and 12 macroimage configurations across three lens systems. Intrinsic variability and observational noise are omitted to isolate the effects of the microlens populations.

We use ten variability and event metrics with ridge regression to test whether omitted physics is mistaken for a change in disk size. We also compare caustic-crossing rates, high-magnification events, and response-lag distributions. Throughout this work we use a flat $\Lambda$CDM cosmology with $H_0=70\,\mathrm{km\,s^{-1}\,Mpc^{-1}}$ and $\Omega_{\rm m}=0.3$.

Section~\ref{sec:two_star} illustrates evolving caustics in a two-star example. Section~\ref{sec:simulations} defines the controlled populations, and Section~\ref{sec:measurements} presents their measurements and inference results. We discuss their interpretation and limitations in Section~\ref{sec:discussion} and summarize our conclusions in Section~\ref{sec:conclusions}.

\section{An illustrative two-star system}
\label{sec:two_star}

\begin{figure*}
\centering
\includegraphics[width=\textwidth]{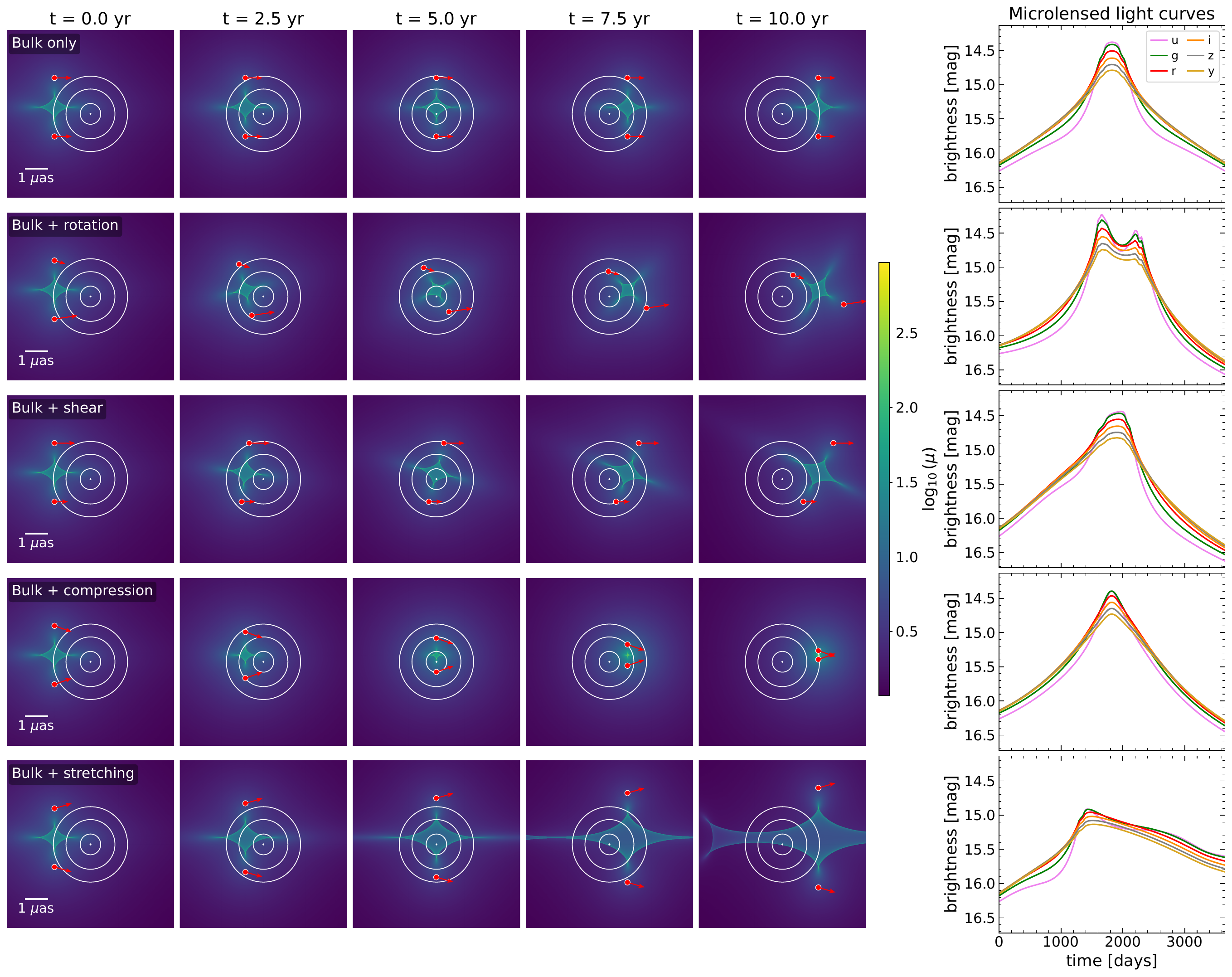}
\caption{Two-star illustration of dynamical microlensing. Each row uses the same initial microlens positions and bulk drift. The five left panels show magnification maps over ten years, and the right panel shows the corresponding six-band $ugrizy$ light curves. The two microlenses each have mass $0.3\,M_\odot$, and the example uses Q2237-like source parameters and redshifts with zero external shear. White contours enclose 68\%, 95\%, and 99\% of the fixed source's $i$-band flux. Red points and arrows show the lens-plane angular positions and velocities, with arrow lengths proportional to speed.}
\label{fig:two_star}
\end{figure*}

Figure~\ref{fig:two_star} illustrates how relative microlens motion reshapes the magnification map and its caustics. With bulk motion alone, the pattern translates without changing shape. Relative motions rotate, shear, compress, or stretch the microlens pair, changing the timing, asymmetry, and relative strengths of the light-curve peaks.

\section{Simulation design}
\label{sec:simulations}

We vary microlens motion and the microlens mass distribution independently, while keeping the source model, mean microlens mass, and distribution of individual microlens velocities fixed. Section~\ref{subsec:lenses_sources} describes the lens configurations and quasar sources, and Section~\ref{subsec:families} defines the four simulation families. Section~\ref{subsec:velocities} explains how we match their velocity distributions, and Section~\ref{subsec:generation} describes the light-curve generation.

\subsection{Lens configurations and quasar sources}
\label{subsec:lenses_sources}

We adopt the lens and quasar configurations for macroimages A--D of Q2237+0305, RX~J1131$-$1231, and HE~0435$-$1223 from Table~1 of \citet{Fagin2026Microcaustics}, including the local convergence, shear, shear orientation, smooth-matter fraction, and relativistic disk geometry. These configurations span different source sizes and local lensing conditions, allowing us to test how the effects vary between systems. Their lens and source redshifts $(z_l,z_s)$ are $(0.0395,1.695)$, $(0.295,0.658)$, and $(0.4546,1.689)$, respectively. Across the 12 macroimages, the convergence, shear, and stellar convergence span $\kappa=0.391$--$0.894$, $\gamma=0.306$--$0.807$, and $\kappa_\star=0.164$--$0.715$, respectively. The stellar convergence is $\kappa_\star=(1-s)\kappa$, where $s$ is the smooth-matter fraction. We infer disk size from each macroimage separately and from all four images jointly. A matched quad comprises four macroimages of the same simulated quasar, sharing the physical observer, lens, and source bulk-velocity draw but with independently sampled local stellar fields.

The source is the relativistic thin disk with lamppost heating implemented in \textsc{Microcaustics}, based on the Novikov--Thorne model \citep{Novikov_1973,PageThorne1974}. Its irradiation and reprocessing conventions follow \citet{chan2024reverberationmappinglamppostwind} and \citet{Fagin2025Reprocessing}. For Q2237, RX~J1131, and HE~0435, respectively, we adopt black-hole masses $\log_{10}(M_{\rm BH}/M_\odot)=(9.08,8.30,8.36)$, Eddington ratios $\lambda_{\rm Edd}=(0.34,0.07,0.19)$, spins $(0.74,0.87,0.70)$, and inclinations $(10^\circ,17^\circ,30^\circ)$. The temperature slope is $\beta=3/4$ and the mean lamp fraction is $f_{\rm lamp}=0.10$. These source parameters are fixed across the four microlensing families. We evaluate the six $ugrizy$ bands at the adopted LSST effective wavelengths of $(3671,4827,6223,7546,8691,9712)\,\text{\AA}$ on a source grid with $1024\times1024$ pixels. The $r$-band elliptical half-light sizes defined in Section~\ref{subsec:inference} satisfy $R_{1/2,r}/\theta_{\rm E}=(0.066,0.074,0.058)$ for Q2237, RX~J1131, and HE~0435, respectively, where $\theta_{\rm E}$ is the angular Einstein radius for $0.3\,M_\odot$.
We omit intrinsic variability by holding the irradiating luminosity fixed at its mean.

\subsection{Four microlensing families}
\label{subsec:families}

We generate four simulation families to isolate the effects of dynamical microlensing and microlens mass distributions:
\begin{enumerate}
\setlength{\itemsep}{0pt}
\setlength{\parsep}{0pt}
\item Independent microlens motions with a mass distribution.
\item Coherent microlens motion with a mass distribution.
\item Independent microlens motions with equal-mass microlenses.
\item Coherent microlens motion with equal-mass microlenses.
\end{enumerate}
Initial positions, sampled masses, and velocities are drawn with independent seeds between families to avoid correlating training and test sets.
The microlens mass distribution is a truncated power law with the Salpeter slope, $p(m)\propto m^{-2.35}$ \citep{Salpeter1955}. We adopt $m_{\max}/m_{\min}=100$ and number-weighted mean $\langle m\rangle=0.3\,M_\odot$, giving mass limits of approximately $0.097$--$9.7\,M_\odot$ \citep{Kochanek_2004,Chan2021MicrolensingDelays}. The single-mass controls set every microlens to $0.3\,M_\odot$. This is a controlled test of the mass distribution rather than a complete model of stellar evolution or the remnant population. The stellar and smooth surface mass densities are held fixed at the adopted macroimage values.

\subsection{Matching the velocity populations}
\label{subsec:velocities}

We retain the observer, lens, and source bulk-motion prescription of \citet{Fagin2026Microcaustics}. To isolate relative microlens motion, the coherent control includes a shared random microlens velocity rather than simply removing the velocity dispersion. In lens-plane angular units, the assigned velocities are
\begin{equation}
\boldsymbol\omega_k^{\rm dyn}=\boldsymbol\omega_{\rm bulk}+\boldsymbol u_k,
\qquad
\boldsymbol\omega_k^{\rm coh}=\boldsymbol\omega_{\rm bulk}+\boldsymbol u,
\label{eq:velocities}
\end{equation}
where $k$ indexes the microlenses. Each $\boldsymbol u_k$ is an independent two-dimensional Gaussian draw with zero mean and component dispersion $\sigma_\omega$, while $\boldsymbol u$ is drawn independently once per macroimage realization from the same distribution and shared by every microlens in that macroimage. The bulk drift combines independent Gaussian peculiar velocities of the lens and source with the fixed projected observer motion, following \citet{Fagin2026Microcaustics}. Averaging over these draws at fixed bulk drift, both families have the same single-microlens velocity distribution and RMS angular speed,
\begin{equation}
\omega_{\rm rms}^2
=|\boldsymbol\omega_{\rm bulk}|^2+2\sigma_\omega^2.
\label{eq:velocity_moment}
\end{equation}

For a population RMS bulk angular speed $\omega_{\rm bulk,rms}$, the theoretical ratio $\sqrt{2}\sigma_\omega/\omega_{\rm bulk,rms}$ is $0.697$, $1.066$, and $0.660$ for Q2237+0305, RX~J1131$-$1231, and HE~0435$-$1223, respectively. These ratios follow from the adopted Gaussian peculiar-velocity dispersions and fixed projected observer motion, and are the same for all images of each system. The bulk term excludes the additional shared random velocity in the coherent control. These ratios show that random microlens speeds are comparable to the bulk speed in all three systems.

\subsection{Light-curve generation}
\label{subsec:generation}

We generate 2000 matched quads per family and system, giving 96,000 six-band macroimage light curves. Each spans 3650 days sampled at 25-day intervals, giving 147 magnification epochs. We use inverse polygon mapping \citep{Mediavilla2006} in \textsc{Microcaustics}~\citep{Fagin2026Microcaustics} with nominal settings of $10^7$ image-plane samples, scout sampling ratio $k=2$, lens-equation refinement $r=2$, and interpolated polygon refinement $v=4$. Caustic labels use an $8192\times8192$ determinant grid within the scouted region. The lens-plane aperture, stellar-boundary prescription, and velocity-coordinate transformations follow the same simulator. These numerical settings are common to all four families, and the source remains fixed on the projected grid while the microlenses move. The solver, source transfer, and numerical validation are described by \citet{Fagin2026Microcaustics}.

\section{Population analysis and results}
\label{sec:measurements}

We compare caustic-crossing rates and HME rates and durations in Section~\ref{subsec:events}, then describe our disk-size inference method in Section~\ref{subsec:inference}. Section~\ref{subsec:metric_shifts} examines individual-metric alignments and single-image inference, and Section~\ref{subsec:results_systems} presents disk-size estimates using all four images together and their dependence on population size. Section~\ref{subsec:robustness} examines robustness to metric choice, macroimage selection, and monitoring duration. Finally, Section~\ref{subsec:lags} examines microlensing-induced response lags and how their distributions change with monitoring duration.

\subsection{Caustic crossings and high-magnification events}
\label{subsec:events}

Microlens dynamics produce more frequent caustic crossings and more frequent, shorter photometric HMEs when events are pooled across each system's four images. We identify caustic crossings from changes in the binary labels described by \citet{Fagin2026Microcaustics}. We identify HMEs using SciPy's peak-finding routine \citep{Scipy} to locate brightening peaks with a prominence of at least $0.30\,\mathrm{mag}$ and a minimum separation of 50 days, using the same rule in each band and data set. Prominence measures a brightening peak's height above the higher of its two surrounding minima, identified by SciPy's peak-prominence calculation. The magnitude threshold matches that adopted by \citet{Neira2025HME}, although our peak-based event definition differs. We measure event durations as the full width at half prominence and exclude peaks at the light-curve endpoints. A caustic crossing can produce a weak photometric feature after finite-source averaging, while a prominent brightening can occur without a caustic passing over the source center.

\label{subsec:results_events}

\begin{figure*}
\centering
\includegraphics[width=\textwidth]{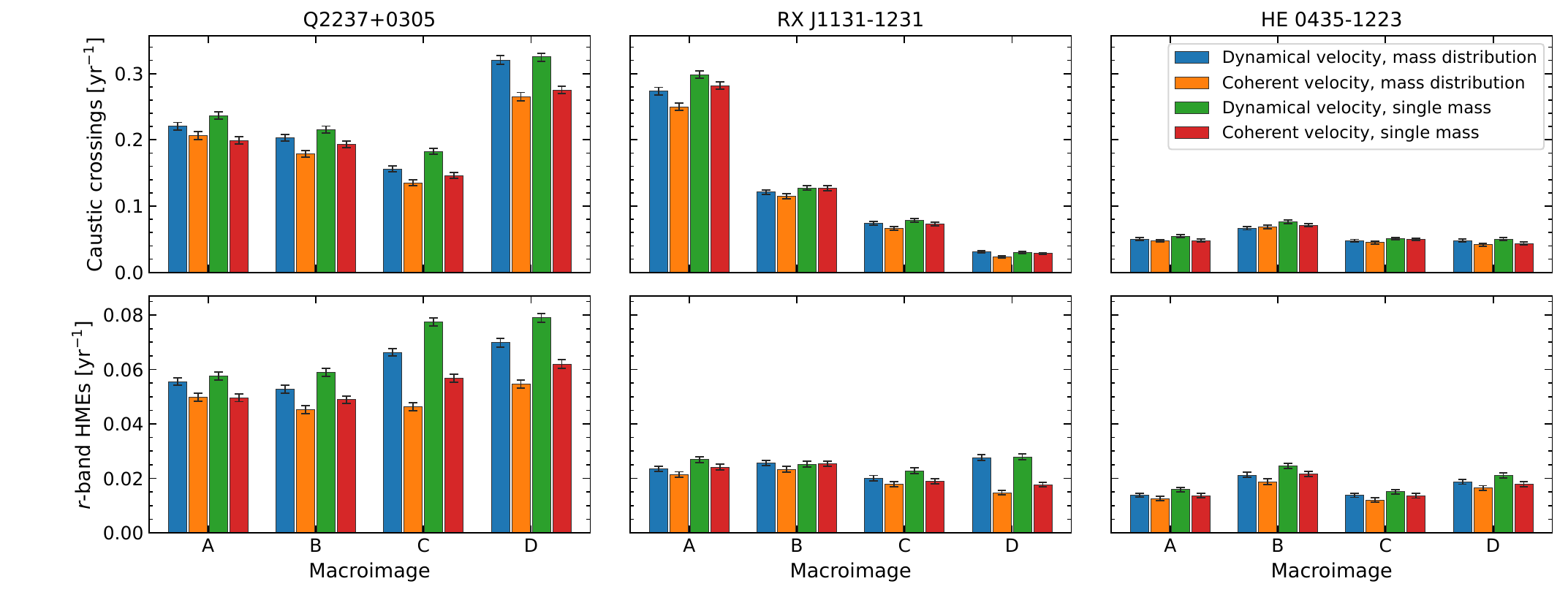}
\caption{Caustic-crossing rates (top) and $r$-band HME rates (bottom) for the four microlensing families. Error bars show central 68\% intervals from 1000 bootstrap resamples of whole quads, drawn independently within each data set. Each family and system contains 2000 matched quads.}
\label{fig:event_rates}
\end{figure*}

Figure~\ref{fig:event_rates} compares caustic-crossing and $r$-band HME rates for the four data sets. Microlens dynamics increase the rates summed across all four images by approximately 15\%, 10\%, and 5\% for caustic crossings and 25\%, 25\%, and 13\% for HMEs in Q2237, RX~J1131, and HE~0435, respectively. Across individual images, dynamic-to-coherent rate ratios range from 0.97 to 1.34 for caustic crossings and from 1.10 to 1.87 for HMEs.

Pooling the detected $r$-band HMEs across all 2000 realizations and the four images of each system, the dynamical simulations with a mass distribution have median durations of 493, 697, and 833 days for Q2237, RX~J1131, and HE~0435, respectively. These are approximately 12\%, 10\%, and 4\% shorter than in the corresponding coherent-motion simulations. Median event prominences change much less, decreasing by only 3--4\%. Microlens dynamics therefore produce more frequent and shorter events without a comparable change in their typical prominence.

The adopted mass distribution reduces caustic-crossing rates by 6\%, 7\%, and 8\% and $r$-band HME rates by 11\%, 6\%, and 12\% relative to the dynamical single-mass simulations for Q2237, RX~J1131, and HE~0435, respectively. Median event durations differ by less than 2\% and median prominences by less than 4\%. For these simulations and event criteria, the mass distribution therefore affects the frequency of detected events more than their typical duration or prominence.

\subsection{Inferring disk size from light-curve metrics}
\label{subsec:inference}

We learn how the light-curve statistics change with disk size in each simplified simulation family. We use ridge regression \citep{HoerlKennard1970} because it gives a simple linear relation between the input statistics and logarithmic disk size, with a penalty that helps limit overfitting when the inputs are correlated. We then apply the fitted relation from each family to the target simulations with dynamical microlensing and a microlens mass distribution. The additional size error, relative to recovery within the training family, measures the effect of the simplified physical assumptions.

The half-light radius is a useful source-size measure because magnification distributions for circular sources depend mainly on this radius rather than the detailed radial profile \citep{Mortonson2005}. For each band $b$, we define $R_{1/2,b}$ as the semimajor axis of an elliptical aperture containing half the unlensed disk light,
\begin{equation}
\int_{\rho_{\rm ell}\leq R_{1/2,b}} I_{\nu,b}(\boldsymbol r)\,d^2\boldsymbol r
=\frac{1}{2}\int I_{\nu,b}(\boldsymbol r)\,d^2\boldsymbol r,
\label{eq:half_light}
\end{equation}
where $\rho_{\rm ell}^2=x'^2+y'^2/\cos^2 i$, $i$ is the adopted disk inclination, and $(x',y')$ are projected source-plane coordinates centered on the disk and aligned with its major and minor axes. We integrate the observed specific intensity $I_{\nu,b}$, evaluated at band $b$'s effective wavelength and including relativistic transfer and inclination. The ellipse defines the flux aperture, not an approximation to the brightness image or an exact relativistic deprojection. The six wavelength-dependent source profiles supply the size range for training.

\begin{table*}
\centering
\caption{The ten light-curve metrics used for population inference.}
\label{tab:metrics}
\begin{tabular}{llp{0.51\textwidth}}
\hline\hline
Metric & Units & Definition \\
\hline
Brightness standard deviation & mag & Temporal standard deviation of $m(t)$. \\
Raw range & mag & $\max m-\min m$. \\
Detrended range & mag & Range after subtracting the best-fitting linear trend. \\
First-derivative tail & dimensionless & $\log_{10}(D_{1,95}/[\mathrm{mag\,day}^{-1}])$. \\
Second-derivative tail & dimensionless & $\log_{10}(D_{2,95}/[\mathrm{mag\,day}^{-2}])$. \\
Third-derivative tail & dimensionless & $\log_{10}(D_{3,95}/[\mathrm{mag\,day}^{-3}])$. \\
\hline
HME rate & yr$^{-1}$ & Number of detected brightening events per monitored year. \\
Clean-HME rate & yr$^{-1}$ & Event rate for the clean subset defined in the text. \\
HME duration & day & Event width at half prominence. \\
Clean-HME duration & day & Half-prominence width for the clean subset. \\
\hline
\end{tabular}
\end{table*}

We summarize the light curves using six continuous-variability metrics and four HME metrics, listed in Table~\ref{tab:metrics}. For the derivative metrics, we use $D_{n,95}=q_{95}(|d^n m/dt^n|)$, where $q_{95}$ denotes the 95th percentile and derivatives are evaluated successively using second-order finite differences. All ten metrics are insensitive to constant magnitude offsets. The HME definition is given in Section~\ref{subsec:events}. We define clean HMEs as brightening events containing one or two sampled caustic crossings within their half-prominence width. These events are particularly useful for probing inner-disk structure because their profiles are less complicated by overlapping caustic crossings.

We summarize each metric by its mean and lower and upper scatter about that mean, retaining information about asymmetric spread and strong excursions across the population. For $N$ values $x_j$ with mean $\bar x$, the two scatters are
\begin{equation}
s_\pm^2=\frac{1}{N}\sum_{j=1}^{N}\max\!\left[\pm(x_j-\bar x),0\right]^2.
\label{eq:population_scatter}
\end{equation}
Both use the full count, so $s_-^2+s_+^2$ equals the population variance. The ten metrics therefore supply 30 input statistics, or features, per population, band, and macroimage. We summarize variability metrics and event rates across light curves. For durations, we combine the detected events from those curves and replace $N$ with the total number of events. A population with no detected events has missing durations rather than zero durations. Undefined duration statistics are filled with their medians in the training simulations, and statistics never observed in training are excluded.

Before taking logarithms, we impose a numerical floor on each derivative metric to avoid learning from numerical artifacts of the simulator. We calculate each metric for Gaussian noise sequences with a standard deviation of 0.1 mmag, using the same sampling and derivative calculation as the light curves, and take the median result as its floor. We exclude a metric's population mean if it lies at the floor in at least 99\% of training examples, keeping this exclusion fixed for validation and prediction.

Ridge regression minimizes the sum of squared prediction residuals plus a penalty on the squared coefficients, leaving the intercept unpenalized. The regression predicts $\log_{10}(R_{1/2,b}/\mu\mathrm{as})$ from the 30 statistics, each standardized by subtracting its training mean and dividing by its training standard deviation. This approach is similar to \citet{JimenezVicenteMediavilla2025ML}, but our inputs summarize populations of light curves rather than individual curves.

Our main disk-size inference comparisons use a target population with dynamical microlensing and a mass distribution, and models trained on three simplified data sets. Coherent motion with a mass distribution neglects relative microlens motion, dynamic motion with a single mass neglects the mass distribution, and coherent motion with a single mass neglects both. Each fitted relation is applied to the target without refitting. Test realizations from its own family measure recovery under the same physical assumptions as the training data, which we call the same-physics test. We define the signed inference error as
\begin{equation}
\Delta\log_{10}R_{1/2,b}
=\log_{10}\!\left(\frac{\widehat R_{1/2,b}}{R_{1/2,b}}\right),
\label{eq:radius_error}
\end{equation}
where $\widehat R_{1/2,b}$ is the inferred radius and $R_{1/2,b}$ is the true simulated radius, so positive values indicate an overestimate.

We fit models using population light-curve statistics from each macroimage separately and from all four jointly. Each individual-image model uses 30 input statistics per band. The joint model combines the four images' statistics into 120 inputs and predicts their common source radius. The two analyses use the same matched population draws and cross-validation folds, but each has its own preprocessing and selected penalty. \citet{JimenezVicenteMediavilla2025ML} similarly combined features from three differential light curves to infer a common disk size. Here we combine population statistics from four images rather than average four separately inferred sizes.

We use populations of $N=128$ matched quads for the main results and examine $N=1,2,4,8,16,32,64,$ and $128$. Each fit uses 10,000 training populations and 2000 populations from each validation or test subset, formed by drawing $N$ quads with replacement. Draws are independent between physical families, with each quad's four macroimages and six bands kept together. Each population provides one example per band, covering the six sampled radii. Joint models use all four macroimages, while individual-image models use the corresponding $N$ curves. These draws reuse the existing simulations rather than generate new independent realizations.

We use five-fold cross-validation, holding out 400 of the 2000 quads in each fold. Within each training subset, three-fold cross-validation compares ridge penalties $(0.1,1,10,100,1000)$. For each system, training family, and $N$, we select one common penalty used in all five outer fits by minimizing the mean inner-validation root-mean-square error (RMSE) across the five outer partitions. Individual-image models select their own common penalties in the same way. Input preparation uses only each fit's training subset, and the target family is partitioned separately. Target predictions and outer-test residuals are not used to select the penalty, but pooling selection across folds means the same-physics test errors are not fully independent of that selection.

For each sampled population, we take the median logarithmic radius error in Equation~\eqref{eq:radius_error} across its six bands. For each simplified model, we subtract the errors on independently drawn training-family test populations from the target errors and take the median of these differences within each outer fold. This baseline correction accounts for offsets that occur even when the training and test simulations use the same physics. For comparison, we train and test a control on independent simulations from the same dynamic population with a mass distribution, using the same folds, input summaries, and penalty-selection procedure. We do not apply the baseline correction to this control. We report the median of the five fold estimates. We calculate RMSE from the per-band prediction errors on test populations from the training family, pooling all five folds.

We estimate finite-sample uncertainty at $N=128$, including for the control, using 100 whole-quad bootstrap resamples drawn independently within each physical family. Each resample repeats input preparation, penalty selection, and all five cross-validation fits, retaining every quad's original outer- and inner-fold assignments. We keep the original central estimates and use the bootstrap distribution of median fold offsets for the uncertainty intervals. We convert offsets and interval endpoints to percentage radius changes using $100(10^\Delta-1)$, where $\Delta$ is the baseline-corrected log-radius offset.

\subsection{Individual metrics and single-image inference}
\label{subsec:metric_shifts}

\begin{figure*}[tp]
\centering
\includegraphics[width=\textwidth]{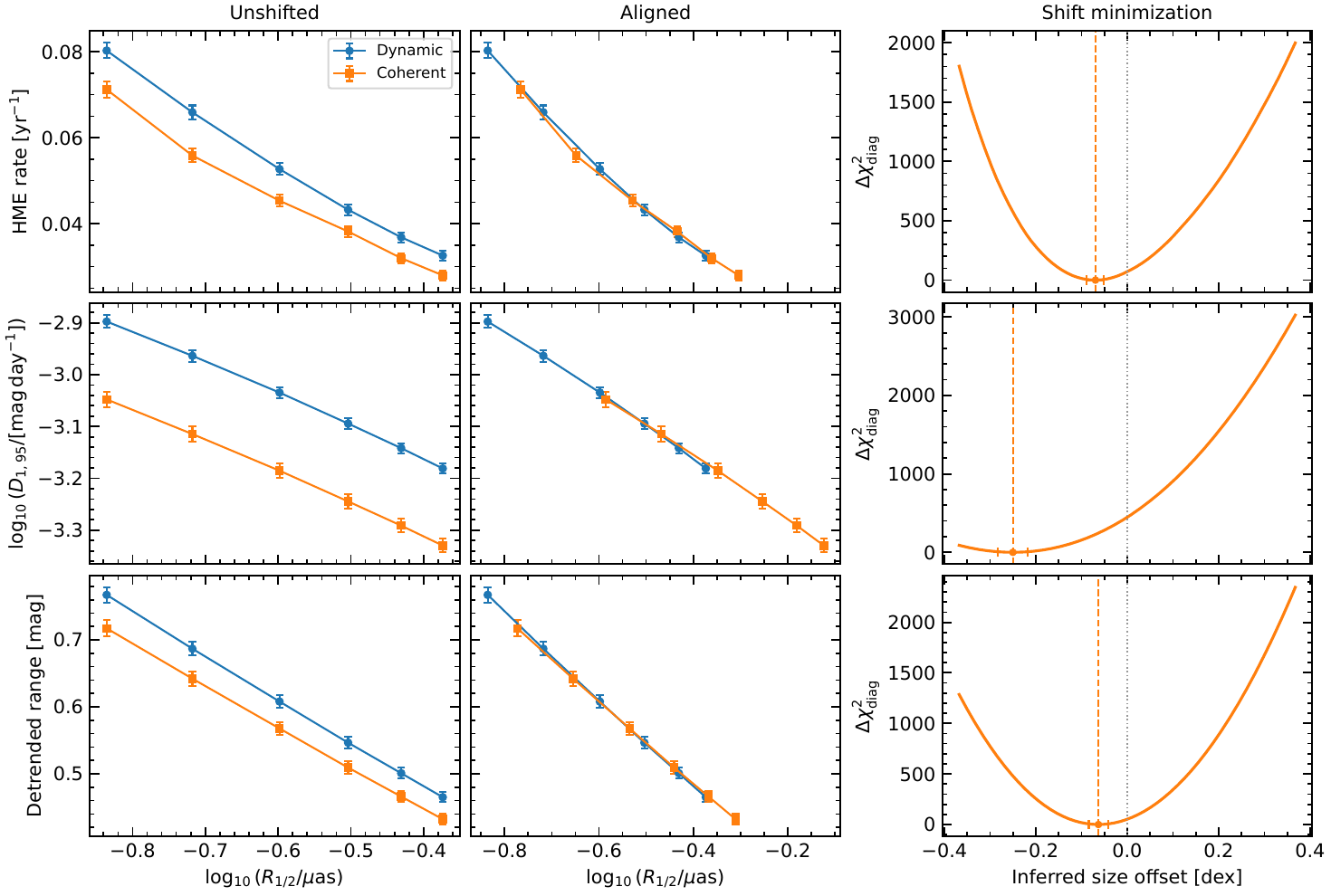}
\caption{Example metric alignment between dynamical and coherent simulations with a mass distribution for Q2237 image B, using HME rate, the first-derivative tail, and detrended range. Columns show the original curves, their best alignment, and the weighted mismatch above its minimum. The right column uses inferred size offset $-\delta$, with dashed lines marking the estimates. Error bars show central 68\% intervals from 1000 bootstrap resamples.}
\label{fig:metric_alignment}
\end{figure*}

We first examine whether changing the microlensing physics shifts individual metrics in the same way as changing disk size. These alignments help interpret the variability changes. Figure~\ref{fig:metric_alignment} illustrates the individual-metric comparison for Q2237 image B.

Let $M_{\rm target}(x)$ and $M_{\rm cal}(x)$ denote the target and comparison population summaries at $x=\log_{10}(R_{1/2}/\mu\mathrm{as})$. We use the population mean of each of the six variability metrics, the population rate for HME counts, and the mean pooled detected-event duration for HME durations. These choices are fixed across all macroimages and comparisons. The individual-metric fits use all 2000 curves per family, whereas the regression predictions use test folds. We shift the comparison curve horizontally by $\delta$ and minimize
\begin{equation}
\chi^2_{\rm diag}(\delta)=\sum_{b\in\{u,g,r,i,z,y\}}
\frac{[M_{\rm target}(x_b)-M_{\rm cal}(x_b-\delta)]^2}{s_b^2},
\label{eq:metric_shift}
\end{equation}
where $s_b$ combines the two summaries' bootstrap uncertainties in quadrature, each taken as half its 16th--84th percentile width and held fixed during alignment. A positive horizontal shift corresponds to a negative inferred size offset, $\Delta\log_{10}R_{1/2}=-\delta$. Because the bands are correlated, we use this quantity only to find the best alignment, not to assign a goodness-of-fit probability. We estimate alignment uncertainties by resampling whole light curves independently within each family, preserving their bands and detected events. We search $-1\leq\delta\leq1$ dex using all six bands, with linear interpolation and at most 1 dex of linear extrapolation beyond either endpoint. The intervals are conditional on this interpolation model and do not quantify uncertainty in extrapolating the wavelength-dependent size relation.

Where a metric depends only weakly on source radius, small population differences can produce large, unstable shifts. We flag poorly constrained metric alignments, including weak size trends, unstable fits, excessive extrapolation, and shifts outside $\pm0.4$ dex. We calculate a descriptive arithmetic mean of the unflagged shifts, averaging within each shared bootstrap draw to preserve correlations between metrics. The contributing metrics can differ between comparison families and macroimages, and the uncertainty intervals are conditional on this selection. Neither these shifts nor their flags enter the ridge regression.

\begin{figure*}[tp]
\centering
\includegraphics[width=0.985\textwidth]{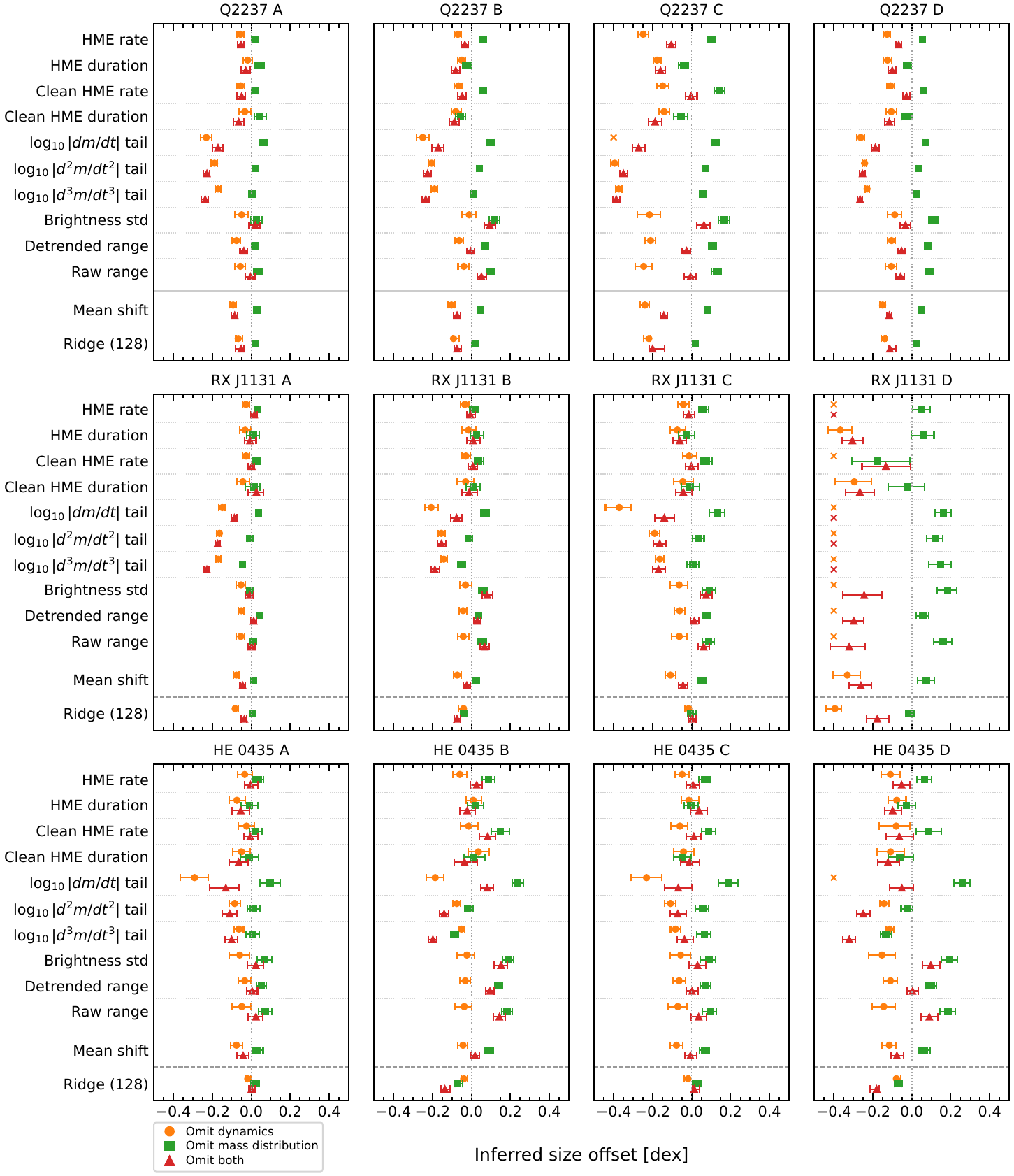}
\caption{Individual metric shifts, their unflagged mean, and the separate ridge estimate at $N=128$. Colors identify the three simplified microlensing assumptions compared with the dynamic target with a mass distribution. Error bars show central 68\% intervals from 1000 bootstrap resamples for the individual metrics and their mean, and 100 complete-analysis bootstrap resamples for ridge. Crosses mark failed or out-of-range individual-metric fits. Flagged central offsets beyond $\pm0.4$ dex are displayed at the cutoff. Negative offsets indicate underestimated radii.}
\label{fig:metric_shift_summary}
\end{figure*}

Figure~\ref{fig:metric_shift_summary} compares the individual metric shifts across all 12 macroimages. Similar shifts suggest that different metrics favor a common size offset, while disagreement indicates that the inferred offset depends on the statistic used. The final row gives the separate ridge estimate at $N=128$, which uses the population means and lower and upper scatters of all ten metrics rather than their fitted shifts.

The central motion-related ridge offset is negative in all twelve macroimages, ranging from $-0.39$ to $-0.015$ dex. The $-0.39$ dex offset for RX~J1131 D is sensitive to extrapolation, with about 68\% of its target predictions outside the training radius range, taking the median across folds. The baseline correction does not correct extrapolation, so large inferred shifts must be assessed alongside recovery within the training family and the sampled radius range. The offsets from omitting the mass distribution span $-0.070$ to $+0.024$ dex and change sign between images. These differences show why a single radius correction cannot describe all images and model assumptions.

\subsection{Joint inference from the four images}
\label{subsec:results_systems}

\begin{table*}
\centering
\caption{Individual-image and joint matched-quad ridge inference at $N=128$. Radius offsets and same-physics test RMSE are in dex. Offset uncertainties are central 68\% bootstrap intervals. The omission offsets include the baseline correction.}
\label{tab:system_inference}
\small
\setlength{\tabcolsep}{2.5pt}
\begin{tabular}{ll@{\hspace{3pt}{\color[gray]{0.8}\vrule width 0.3pt}\hspace{3pt}}rr@{\hspace{3pt}{\color[gray]{0.8}\vrule width 0.3pt}\hspace{3pt}}rr@{\hspace{3pt}{\color[gray]{0.8}\vrule width 0.3pt}\hspace{3pt}}rr@{\hspace{3pt}{\color[gray]{0.8}\vrule width 0.3pt}\hspace{3pt}}rr}
\hline\hline
 & & \multicolumn{2}{c@{\hspace{3pt}{\color[gray]{0.8}\vrule width 0.3pt}\hspace{3pt}}}{Omit dynamics} & \multicolumn{2}{c@{\hspace{3pt}{\color[gray]{0.8}\vrule width 0.3pt}\hspace{3pt}}}{Omit mass distribution} & \multicolumn{2}{c@{\hspace{3pt}{\color[gray]{0.8}\vrule width 0.3pt}\hspace{3pt}}}{Omit both} & \multicolumn{2}{c}{Control} \\
\cline{3-4}\cline{5-6}\cline{7-8}\cline{9-10}
System & Images & Offset & RMSE & Offset & RMSE & Offset & RMSE & Offset & RMSE \\
\hline
Q2237+0305 & A & $-0.067^{+0.021}_{-0.012}$ & $0.033$ & $+0.022^{+0.005}_{-0.013}$ & $0.024$ & $-0.052^{+0.013}_{-0.029}$ & $0.032$ & $-0.005^{+0.011}_{-0.001}$ & $0.028$ \\
 & B & $-0.092^{+0.028}_{-0.003}$ & $0.028$ & $+0.017^{+0.012}_{-0.016}$ & $0.032$ & $-0.075^{+0.023}_{-0.015}$ & $0.031$ & $+0.003^{+0.006}_{-0.011}$ & $0.030$ \\
 & C & $-0.219^{+0.009}_{-0.027}$ & $0.030$ & $+0.019^{+0.012}_{-0.013}$ & $0.031$ & $-0.200^{+0.063}_{-0.017}$ & $0.033$ & $+0.001^{+0.006}_{-0.005}$ & $0.027$ \\
 & D & $-0.140^{+0.008}_{-0.017}$ & $0.031$ & $+0.024^{+0.013}_{-0.006}$ & $0.025$ & $-0.113^{+0.031}_{-0.003}$ & $0.027$ & $+0.005^{+0.002}_{-0.011}$ & $0.026$ \\
 & A--D jointly & $-0.142^{+0.030}_{-0.004}$ & $0.019$ & $+0.022^{+0.005}_{-0.011}$ & $0.023$ & $-0.109^{+0.040}_{-0.008}$ & $0.020$ & $-0.002^{+0.006}_{-0.002}$ & $0.016$ \\
\hline
RX~J1131$-$1231 & A & $-0.083^{+0.015}_{-0.009}$ & $0.030$ & $+0.007^{+0.012}_{-0.015}$ & $0.028$ & $-0.035^{+0.013}_{-0.017}$ & $0.032$ & $-0.003^{+0.004}_{-0.003}$ & $0.024$ \\
 & B & $-0.040^{+0.006}_{-0.026}$ & $0.037$ & $-0.043^{+0.018}_{-0.010}$ & $0.029$ & $-0.074^{+0.019}_{-0.016}$ & $0.032$ & $-0.010^{+0.013}_{-0.000}$ & $0.030$ \\
 & C & $-0.015^{+0.006}_{-0.019}$ & $0.032$ & $-0.006^{+0.028}_{-0.001}$ & $0.037$ & $+0.002^{+0.021}_{-0.020}$ & $0.043$ & $+0.003^{+0.007}_{-0.008}$ & $0.032$ \\
 & D & $-0.394^{+0.035}_{-0.045}$ & $0.047$ & $-0.007^{+0.022}_{-0.022}$ & $0.059$ & $-0.177^{+0.061}_{-0.054}$ & $0.060$ & $+0.003^{+0.009}_{-0.007}$ & $0.043$ \\
 & A--D jointly & $-0.102^{+0.019}_{-0.002}$ & $0.020$ & $-0.032^{+0.016}_{-0.007}$ & $0.020$ & $-0.056^{+0.017}_{-0.009}$ & $0.023$ & $-0.003^{+0.005}_{-0.004}$ & $0.018$ \\
\hline
HE~0435$-$1223 & A & $-0.018^{+0.017}_{-0.007}$ & $0.040$ & $+0.017^{+0.025}_{-0.018}$ & $0.043$ & $+0.001^{+0.020}_{-0.012}$ & $0.044$ & $+0.003^{+0.009}_{-0.005}$ & $0.032$ \\
 & B & $-0.039^{+0.019}_{-0.014}$ & $0.039$ & $-0.070^{+0.025}_{-0.013}$ & $0.035$ & $-0.136^{+0.028}_{-0.020}$ & $0.034$ & $-0.004^{+0.014}_{-0.006}$ & $0.040$ \\
 & C & $-0.018^{+0.007}_{-0.019}$ & $0.041$ & $+0.021^{+0.028}_{-0.007}$ & $0.040$ & $+0.013^{+0.029}_{-0.008}$ & $0.047$ & $+0.005^{+0.007}_{-0.008}$ & $0.036$ \\
 & D & $-0.079^{+0.024}_{-0.006}$ & $0.041$ & $-0.070^{+0.019}_{-0.020}$ & $0.040$ & $-0.181^{+0.014}_{-0.035}$ & $0.040$ & $-0.002^{+0.010}_{-0.006}$ & $0.034$ \\
 & A--D jointly & $-0.025^{+0.015}_{-0.005}$ & $0.024$ & $-0.053^{+0.046}_{-0.005}$ & $0.025$ & $-0.088^{+0.031}_{-0.008}$ & $0.025$ & $-0.001^{+0.008}_{-0.002}$ & $0.023$ \\
\hline
\end{tabular}
\tablecomments{Offset intervals come from 100 whole-quad bootstrap resamples, each repeating preprocessing, penalty selection, and all five cross-validation fits. Central estimates retain the original median fold offsets. RMSE measures recovery on test simulations from each model's training family, combining squared residuals across the five folds.}
\end{table*}

For populations of $N=128$ matched quads, omitting relative microlens motion gives inferred radii $27.8^{+0.6}_{-5.1}\%$, $20.9^{+0.3}_{-3.6}\%$, and $5.6^{+1.2}_{-3.3}\%$ smaller in the joint fits for Q2237, RX~J1131, and HE~0435, respectively. Table~\ref{tab:system_inference} gives both the individual-image and joint offsets in dex.

Omitting the mass distribution gives radius offsets of $+5.2^{+1.3}_{-2.7}\%$, $-7.1^{+3.5}_{-1.4}\%$, and $-11.5^{+9.9}_{-1.0}\%$ for Q2237, RX~J1131, and HE~0435, respectively. Its effect therefore differs in both size and direction between systems. Omitting both ingredients gives $-22.2^{+7.4}_{-1.4}\%$, $-12.1^{+3.6}_{-1.8}\%$, and $-18.4^{+6.0}_{-1.4}\%$. This produces a smaller radius underestimate than omitting microlens motion alone for Q2237 and RX~J1131, but a larger underestimate for HE~0435. Each training model relates the light-curve statistics to disk size differently, so the combined inference offset cannot be treated as the sum of two independent corrections.

Across the nine joint fits, the test RMSE within the training family ranges from $0.019$ to $0.025$ dex. Joint inference preserves the correlations from shared bulk velocities and can give a different estimate from averaging the individual-image results.

When trained and tested on independent simulations that both include microlens dynamics and a mass distribution, the control recovers the joint radii with median offsets of $-0.002$, $-0.003$, and $-0.001$ dex and RMSEs of $0.016$, $0.018$, and $0.023$ dex for Q2237, RX~J1131, and HE~0435, respectively. Its small residual offsets provide a check on recovery when both physical ingredients are included.

An offset is already present when the population contains only one matched quad ($N=1$). The model trained without microlens dynamics gives $-0.068$ dex for Q2237, or approximately $15\%$ smaller radii. RX~J1131 and HE~0435 give central offsets of $-0.041$ and $-0.009$ dex. The corresponding training-family test RMSEs are $0.133$, $0.156$, and $0.148$ dex, larger than these median offsets. A systematic shift can therefore be visible across repeated realizations even when individual estimates have substantial scatter.

\begin{figure*}
\centering
\includegraphics[width=\textwidth]{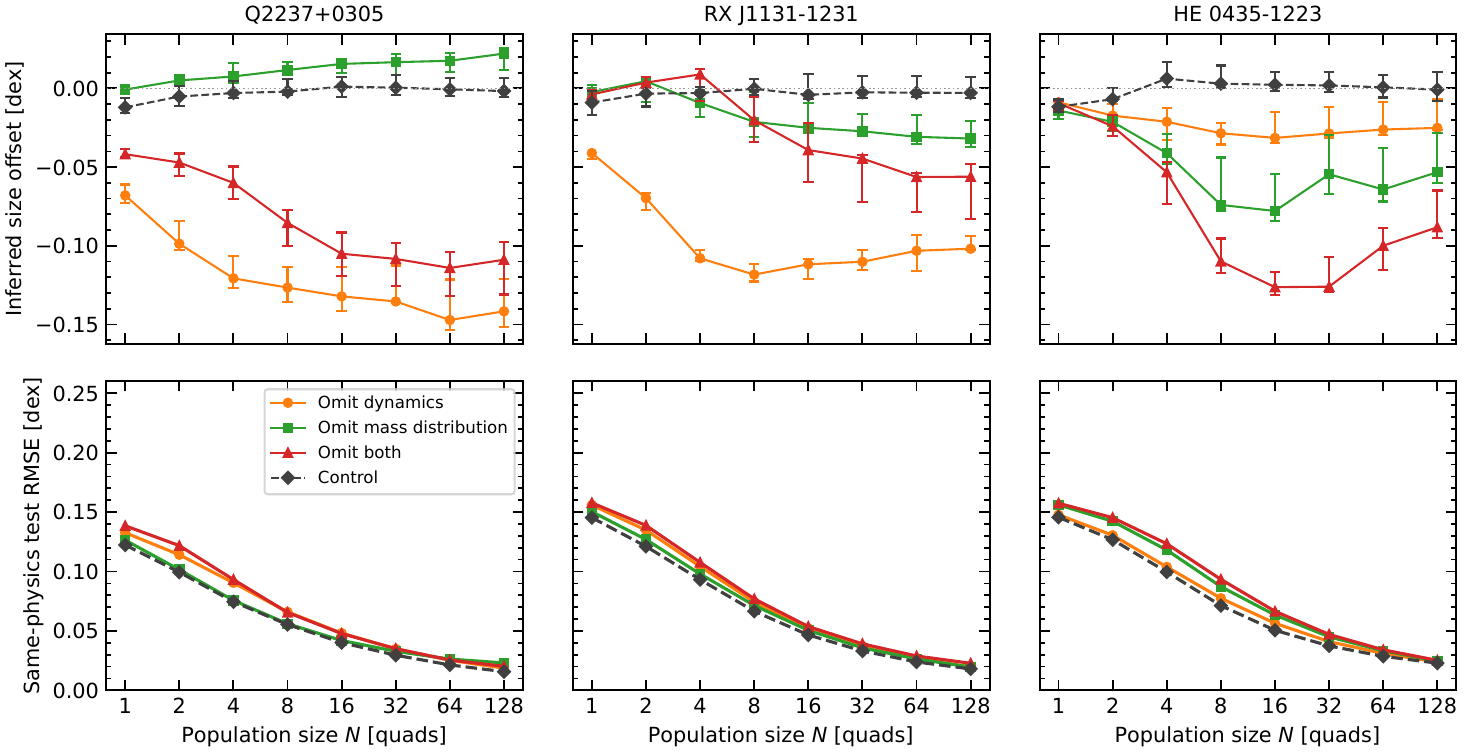}
\caption{Joint ridge size offsets (top) and same-physics test RMSE (bottom) versus population size. The omission results include the baseline correction. No correction is applied to the control, which is trained and tested on independent dynamic simulations with a mass distribution. The top error bars show the 16th--84th percentile spread across cross-validation fits. This variation can persist or increase even as the same-physics test RMSE decreases.}
\label{fig:population_sizes}
\end{figure*}

Figure~\ref{fig:population_sizes} shows how the joint size offsets and same-physics test RMSE depend on population size. Increasing the population size improves estimates of the light-curve statistics, but does not remove errors caused by simplified microlensing physics. The inferred offset can also change as the population summaries and selected ridge penalty change with $N$. We use the same ten metrics, mean and lower/upper scatter summaries, and realization-level splits throughout. 

\subsection{Robustness to metric choice, macroimage selection, and monitoring duration}
\label{subsec:robustness}

Disk-size biases may depend on metric choice, macroimage selection, and monitoring duration. We repeat the $N=128$ analysis excluding individual metrics, metric groups, or one macroimage, and using the first five or seven years. Each test retains the original cross-validation splits and refits the models. Appendix~\ref{app:robustness} gives all the resulting offsets and prediction errors.

Removing both duration metrics or all four HME metrics changes the offsets from neglecting dynamics or the mass distribution by less than 0.9 percentage points. Removing each metric individually preserves the direction of all six biases. The dynamical offsets range from $-29\%$ to $-25\%$, $-23\%$ to $-16\%$, and $-6\%$ to $-5\%$ for Q2237, RX~J1131, and HE~0435, respectively. The corresponding mass-distribution offsets range from $+2\%$ to $+6\%$, $-8\%$ to $-2\%$, and $-16\%$ to $-3\%$. Thus, the directions remain consistent, while the inferred magnitudes depend on metric choice, particularly for the mass distribution.

The three derivative metrics alone give similar prediction accuracy and broadly similar offsets to the full model. Excluding all three increases the same-physics test RMSE from approximately $0.019$--$0.025$ dex to $0.027$--$0.038$ dex and substantially reduces the magnitudes of the offsets. These metrics therefore carry much of the disk-size information and sensitivity to the assumed microlensing physics.

Excluding each macroimage in turn gives dynamical radius underestimates of 23--33\%, 9--30\%, and 4--8\% for Q2237, RX~J1131, and HE~0435, respectively. The corresponding offsets from omitting the mass distribution range from $+5\%$ to $+6\%$, $-8\%$ to $+0.5\%$, and $-14\%$ to $-5\%$. Thus, the dynamical bias remains negative for every three-macroimage combination, although its magnitude, particularly for RX~J1131, depends on which macroimages are included.

The dynamical disk-size bias persists in both shorter light curves, with radius underestimates of 24--26\%, 15--18\%, and 5--6\% for Q2237, RX~J1131, and HE~0435, respectively. The corresponding 68\% bootstrap intervals remain below zero in all three systems. Its larger magnitude in longer Q2237 and RX~J1131 light curves may partly reflect improved disk-size sensitivity. Mass-distribution offsets depend more strongly on monitoring duration. For example, the Q2237 offset changes from approximately $-1\%$ at five years to $+5\%$ at ten years, although its five-year interval includes zero, while RX~J1131 changes from $-17\%$ to $-7\%$.

Microlens dynamics increase caustic-crossing and HME rates at all three monitoring durations. Median HME durations remain approximately 11--14\%, 8--10\%, and 2--4\% shorter for Q2237, RX~J1131, and HE~0435, although the five-year uncertainty interval includes zero for HE~0435. Absolute median durations increase with monitoring baseline, showing that measured widths depend on how much of each event is observed. The reported duration differences therefore apply to events detected within the tested observing periods, and longer simulations would help establish their behavior beyond ten years.

\subsection{Response-lag distributions and monitoring duration}
\label{subsec:lags}

\begin{figure*}
\centering
\includegraphics[width=0.95\textwidth]{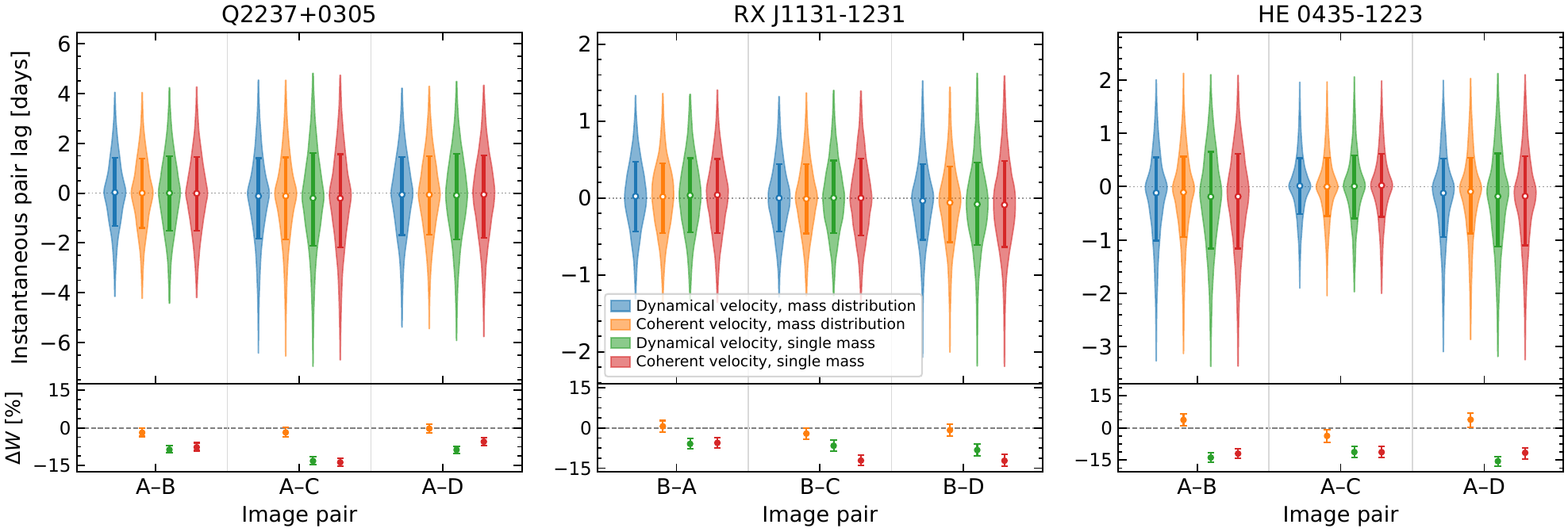}
\caption{Instantaneous $r$-band differential response-lag distributions for the reference-image pairs of each system. Violins show the central 99\%, with medians and 16th--84th percentile population ranges marked. The lower panels show relative-width differences $\Delta W$, with error bars giving central 68\% intervals from 1000 bootstrap resamples of whole quads, drawn independently within each family.}
\label{fig:lag_distributions}
\end{figure*}

\begin{figure*}
\centering
\includegraphics[width=0.95\textwidth]{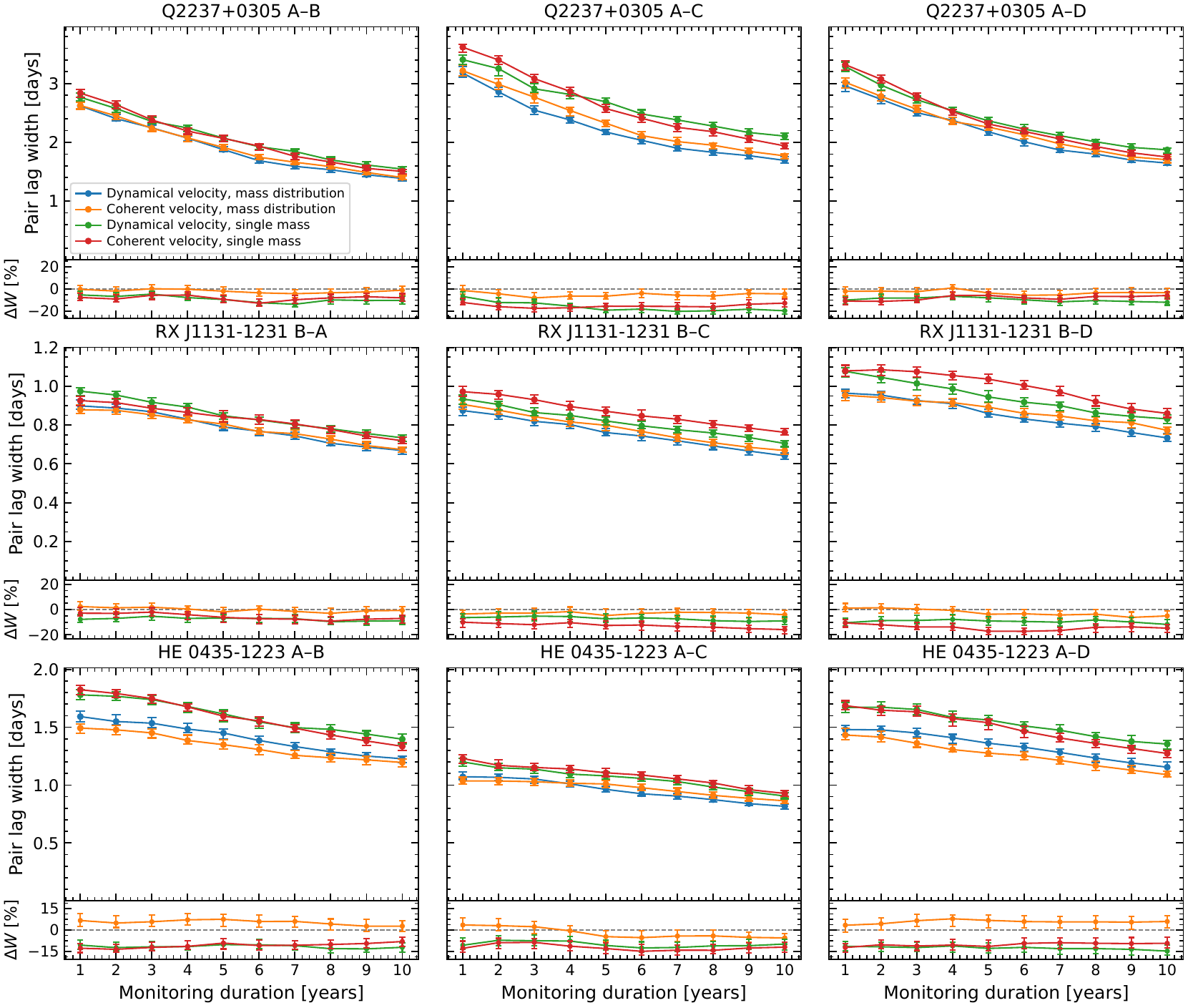}
\caption{$r$-band campaign-averaged differential lag widths $W=P_{84}-P_{16}$ for monitoring durations from 1 to 10 years. Main panels show the absolute widths, while the lower panels show relative-width differences $\Delta W$. Error bars show central 68\% intervals from 1000 bootstrap resamples.}
\label{fig:lag_monitoring}
\end{figure*}

Microlensing weights disk regions with different response delays unequally. We test how microlens dynamics and the mass distribution affect the resulting lag differences between macroimages, both instantaneously and over a monitoring campaign.

For each disk pixel, \textsc{Microcaustics} provides the observed response delay $\tau(\boldsymbol r)$ and linear response weight $W_b(\boldsymbol r)$, evaluated about the mean irradiating luminosity \citep{Fagin2026Microcaustics}. The weight is the change in observed band intensity per unit change in normalized irradiating luminosity. For magnification $\mu_i(\boldsymbol r,t)$ in macroimage $i$, the microlensed mean lag is
\begin{equation}
\overline\tau_{i,b}(t)
=\frac{\int\mu_i(\boldsymbol r,t)W_b(\boldsymbol r)\tau(\boldsymbol r)\,d^2\boldsymbol r}
{\int\mu_i(\boldsymbol r,t)W_b(\boldsymbol r)\,d^2\boldsymbol r}.
\label{eq:mean_lag}
\end{equation}
We report $\delta\tau_{i,b}(t)=\overline\tau_{i,b}(t)-\overline\tau_b^{(0)}$, where $\overline\tau_b^{(0)}$ is obtained by setting $\mu_i=1$ \citep{TieKochanek2018}. We calculate these lags directly from the magnification maps and source response.

Figure~\ref{fig:lag_distributions} shows instantaneous $r$-band lag differences between matched macroimages, pooling all epochs of the 2000 quads per system and family with equal weight per trajectory. We show three pairs per system, referenced to A for Q2237 and HE~0435 and B for RX~J1131. We characterize each distribution by its central 68\% width $W=P_{84}-P_{16}$. The lower panels compare these widths using $\Delta W=100(W_{\rm target}/W_{\rm comparison}-1)$, where the target includes dynamics and a mass distribution. Negative values indicate a narrower target distribution.

With dynamics included, the mass distribution narrows the instantaneous distributions in all nine pairs, with median width reductions of approximately 9\%, 7\%, and 14\% for Q2237, RX~J1131, and HE~0435, respectively. With a mass distribution included, dynamics change the widths by less than 4\% in every pair, with both increases and decreases.

To measure cancellation between positive and negative shifts over longer monitoring periods, we calculate the campaign-averaged differential response lag,
\begin{equation}
\delta t_{ij,b}^{(T)}=\frac{1}{T}\int_0^T
\left[\delta\tau_{i,b}(t)-\delta\tau_{j,b}(t)\right]\,dt,
\label{eq:differential_campaign_lag}
\end{equation}
for durations from one to ten years, starting at the beginning of each trajectory. We use trapezoidal integration with linear interpolation to annual endpoints. Figure~\ref{fig:lag_monitoring} shows the resulting population widths $W$ and their relative differences $\Delta W$.

With dynamics included, the mass distribution narrows the ten-year distributions in all nine pairs, with median reductions of approximately 12\%, 9\%, and 12\% for Q2237, RX~J1131, and HE~0435. Dynamics produce smaller, pair-dependent changes. For the dynamic population with a mass distribution, averaging over ten rather than one year reduces the widths by 22--47\%, leaving ten-year widths of $0.64$--$1.69$ days and a median of 1.2 days across the nine pairs. This narrowing reflects temporal averaging rather than weaker instantaneous perturbations.

\section{Discussion}
\label{sec:discussion}

\subsection{Interpreting population disk-size offsets}
\label{subsec:size_interpretation}

Relative microlens motion changes more than the speed at which a fixed magnification pattern passes the source. Our coherent controls already match the single-microlens velocity distribution, but only independent microlens motions change the separations and orientations of the microlenses. The resulting caustics deform as well as translate, as illustrated in Figure~\ref{fig:two_star}. The increased HME rates and shorter durations are qualitatively consistent with earlier moving-star calculations \citep{KundicWambsganss1993,KundicWittChang1993,WambsganssKundic1995}. A model trained on coherent motion can interpret the additional variability from microlens dynamics as evidence for a smaller disk, which smooths the magnification pattern less. The negative motion-related size offsets are consistent with this interpretation in the tested configurations, but do not separate the effects of caustic deformation and effective speed. The Q2237 offset at $N=1$ shows that the effect is not confined to large populations, although individual predictions have substantially larger scatter.

A mass distribution introduces a range of Einstein-radius scales and changes the magnification pattern that is smoothed by the extended disk. Light-curve statistics can change even when the mean mass is fixed, because finite-source microlensing also depends on the microlens mass range and source size \citep{Schechter2004MassSpectrum,Congdon2007}. The mass-related offsets differ in sign between systems, so they do not support a universal correction.

Matching the single-microlens velocity distribution does not match the speed of the caustic pattern. Published disk-size analyses incorporate microlens motions through effective-velocity prescriptions and marginalize over velocities \citep{Kochanek_2004,Morgan2010}. Our comparison does not determine how much of the inferred offset would remain after optimizing an effective transverse-velocity distribution. A velocity-based correction would need to be tested across lens configurations because the inferred offsets differ between systems.

Microlensing measurements often favor disks larger than predicted from their luminosities under the standard thin-disk model \citep{Morgan2010,Cornachione_2020}. If including microlens motions also increases inferred sizes in observed systems, it would enlarge rather than resolve this discrepancy. \citet{Poindexter10Methods} likewise noted that static patterns could underestimate source sizes by missing faster caustic motion. The magnitude of this effect must be established for the relevant velocity priors, source models, observations, and inference method rather than applied directly to published measurements.

\subsection{Implications for time-delay cosmography}
\label{subsec:lag_cosmography}

Microlens dynamics change the temporal evolution of lag differences but have little effect on their widths in our simulations. Longer monitoring narrows both dynamic and coherent distributions as positive and negative shifts partly cancel. At fixed mean mass, the adopted mass distribution produces narrower lag distributions than equal-mass microlenses, including after ten-year averaging. Earlier calculations also found sensitivity to the microlens population \citep{Chan2021MicrolensingDelays}.

These response delays can perturb inter-image time-delay measurements used in cosmography \citep{TieKochanek2018,Chen2018MicrolensingDelays}. At fixed lens model, a small fitted delay error $\delta t_{ij}$ propagates approximately as
\begin{equation}
\frac{\widehat H_0-H_0}{H_0}\simeq
-\frac{\delta t_{ij}}{\Delta t_{ij}^{\rm cosmo}},
\label{eq:hubble_delay_perturbation}
\end{equation}
where $\Delta t_{ij}^{\rm cosmo}$ is the true cosmological delay. A given error therefore matters more for shorter cosmological delays. Our campaign-averaged response lags characterize a possible contribution to this error, rather than a recovered delay bias.

Microlensing changes both the brightness and temporal response of each macroimage, so their light curves cannot generally be related by a constant time shift \citep{TieKochanek2018}. Small changes in response-lag widths therefore do not establish that dynamics are unimportant for delay recovery. Testing this requires intrinsically variable light curves with realistic cadence and noise, which we leave for future work.

\subsection{Limitations}
\label{subsec:limitations}

This is a controlled simulation experiment, not an inference from observed quasars. We use three lens systems and hold their macro lens and disk parameters fixed. Their macroimages sample different local lensing conditions but are not independent lens galaxies. The measured offsets therefore describe these configurations and inference models, rather than a representative quasar population. Using wavelength to vary the emitting size lets us keep the source field, lens region, and spatial resolution fixed across the comparison. However, wavelength also changes the surface-brightness profile, so these comparisons do not isolate a pure rescaling of the disk.

We omit intrinsic variability, photometric noise, observing gaps, and survey selection to isolate microlens dynamics and the microlens mass distribution. The event rates and durations depend on the adopted source sizes, HME definition, sampling cadence, and monitoring baseline. The checks in Section~\ref{subsec:robustness} show that the dynamical disk-size bias persists without HME metrics, after excluding any one macroimage, and over shorter monitoring baselines, but do not establish convergence of event durations beyond ten years or test these observational effects. Intrinsic variability can obscure microlensing, but also supplies an independent disk-size signal through continuum reverberation \citep{Fagin2025Reprocessing,chan2024reverberationmappinglamppostwind}. Time-delay-aligned difference light curves remove the common intrinsic component and are widely used to isolate microlensing \citep{Kochanek_2004,JimenezVicenteMediavilla2025ML}. Differential magnification of the reverberating continuum and broad-line region prevents exact cancellation and can leave reverberation residuals \citep{Paic2022}. Delay uncertainty and shared measurements can introduce further correlated residuals. Our microlensing-only curves do not include these effects and are not equivalent to noisy observed difference curves.

\subsection{Future work}
\label{subsec:future_work}

The speed of \textsc{Microcaustics} enables larger dynamical training sets spanning realistic distributions of quasars, lens galaxies, and microlens masses \citep{Fagin2026Microcaustics}. These could test how source and velocity uncertainties affect the inferred biases. Including survey cadence, noise, selection, and intrinsic variability would support training nonlinear models, such as deep neural networks, for inference from observed multiband light curves \citep{Fagin2025Reprocessing}.

Similar model dependence may affect inferred sizes and inter-image delays of strongly lensed transients such as supernovae, whose expanding photospheres sample the microlensing pattern over time \citep{DoblerKeeton2006,Goldstein2018SN}. Their shorter light curves allow less stellar displacement than decade-long quasar monitoring at the same velocity, but the microlens mass distribution still sets the caustic scales.

A wavelength-dependent size bias could also affect the inferred temperature profile. In the power-law regime, $T(R)\propto R^{-\beta}$ gives $R_{1/2}\propto\lambda^{1/\beta}$, with $\beta=3/4$ for the standard thin disk. This relation is approximate for an irradiated relativistic disk. A wavelength-independent radius bias changes only the normalization, while testing biases in the temperature slope requires simulations that vary the temperature profile directly.

\section{Conclusions}
\label{sec:conclusions}

Neglecting relative microlens motion can bias population disk-size estimates even when the single-microlens velocity distribution is preserved. We tested this with 96,000 ten-year, six-band light curves generated using \textsc{Microcaustics}. Four simulation families vary microlens motion and the mass distribution independently while holding the mean microlens mass fixed. The velocity-matched comparison isolates relative microlens motion, which allows the caustic network to deform.

Models trained on coherent microlens motion underestimate disk sizes in all twelve macroimages when applied to simulations with independent microlens motions, with a mass distribution included in both cases. At $N=128$ matched quads, the joint estimates correspond to radii $27.8\%$, $20.9\%$, and $5.6\%$ smaller for Q2237, RX~J1131, and HE~0435, respectively. Omitting the mass distribution gives system-dependent shifts of $+5.2\%$, $-7.1\%$, and $-11.5\%$, while omitting both gives $-22.2\%$, $-12.1\%$, and $-18.4\%$.

At $N=128$, training and testing with both microlens dynamics and a mass distribution gives little systematic offset and RMSEs of $0.016$--$0.023$ dex, corresponding to roughly 4--5\% in radius. Models omitting dynamics, the mass distribution, or both achieve similarly small RMSEs of $0.019$--$0.025$ dex on test simulations with the same physical assumptions, but develop systematic offsets when applied to simulations with both effects. Accurate recovery under a model's own assumptions therefore does not guarantee robustness to omitted physics.

Pooling the four images of each system, microlens dynamics increase caustic-crossing rates by 5--15\% and $r$-band HME rates by 13--25\%, while shortening median HME durations by 4--12\% when a mass distribution is included. Median prominences change by only 3--4\%. With microlens dynamics included, the adopted mass distribution reduces these rates by 6--8\% and 6--12\%, respectively, while changing median durations by less than 2\%.

Including the microlens mass distribution narrows instantaneous differential-lag widths by a median of 9\%, 7\%, and 14\% across the three macroimage pairs in Q2237, RX~J1131, and HE~0435, respectively. Microlens dynamics change these widths by less than 4\% in every pair, with both increases and decreases. Longer monitoring narrows the campaign-averaged distributions, but their ten-year central 68\% widths remain $0.64$--$1.69$ days across the nine pairs. Quantifying the impact on time-delay cosmography requires applying time-delay inference to simulated observations that include intrinsic variability, microlensing, and realistic cadence and noise.

In our simulations, larger populations improve disk-size inference precision but do not eliminate biases from omitted microlensing physics. Our results therefore motivate dynamical simulations with a microlens mass distribution for population-level disk inference, or explicit validation of any simplified treatment against them. \textsc{Microcaustics} makes these comparisons practical at the scale required for large training sets.

\section*{Data and code availability}

The simulator, \textsc{Microcaustics}, is open-source \citep{Fagin2026Microcaustics} and available at \url{https://github.com/JFagin/microcaustics}. All code, data, and analysis used in this work are available upon reasonable request from the corresponding author.

\begin{acknowledgments}
Support was provided by Schmidt Sciences, LLC, for J.F., J.C., S.M., M.O., L.P.L., and Y.H\@. S.B., H.B., and P.S. are supported by the US Department of Energy Grant DE-SC0026113.
C.S. acknowledges the support of an NSERC Postdoctoral Fellowship and a CITA National Fellowship.

This work made use of ChatGPT and Codex (\mbox{OpenAI}) assistance for code development and manuscript editing. The authors reviewed and take responsibility for all the content of this paper.
\end{acknowledgments}

\software{\textsc{Microcaustics}~\citep{Fagin2026Microcaustics}, \textsc{Caustics}~\citep{Stone2024Caustics}, \textsc{PyTorch}~\citep{Pytorch}, \textsc{Triton}~\citep{Tillet2019Triton}, \textsc{NumPy}~\citep{Numpy}, \textsc{SciPy}~\citep{Scipy}, \textsc{Matplotlib}~\citep{Matplotlib}.}

%\clearpage
\onecolumngrid
\appendix
\section{Robustness checks}
\label{app:robustness}

Table~\ref{tab:robustness} gives the $N=128$ joint-inference results for the metric-selection, macroimage-selection, and monitoring-duration tests discussed in Section~\ref{subsec:robustness}.

\setcounter{table}{0}
\renewcommand{\thetable}{A\arabic{table}}
\renewcommand{\theHtable}{A\arabic{table}}
\begin{table}[!h]
\centering
\caption{Sensitivity of population disk-size inference to metric selection, macroimage selection, and monitoring duration at $N=128$.}
\label{tab:robustness}
\scriptsize
\setlength{\tabcolsep}{0.4pt}
\renewcommand{\arraystretch}{1.35}
\begin{tabular}{@{}l@{\hspace{0.275pt}{\color[gray]{0.85}\vrule width 0.25pt}\hspace{0.275pt}}c@{\hspace{0.275pt}{\color[gray]{0.85}\vrule width 0.25pt}\hspace{0.275pt}}c@{\hspace{0.275pt}{\color[gray]{0.7}\vrule width 0.25pt}\hspace{0.275pt}}c@{\hspace{0.275pt}{\color[gray]{0.85}\vrule width 0.25pt}\hspace{0.275pt}}c@{\hspace{0.275pt}{\color[gray]{0.7}\vrule width 0.25pt}\hspace{0.275pt}}c@{\hspace{0.275pt}{\color[gray]{0.85}\vrule width 0.25pt}\hspace{0.275pt}}c@{}}
\hline
 & \multicolumn{2}{c@{\hspace{0.275pt}{\color[gray]{0.7}\vrule width 0.25pt}\hspace{0.275pt}}}{Q2237+0305} & \multicolumn{2}{c@{\hspace{0.275pt}{\color[gray]{0.7}\vrule width 0.25pt}\hspace{0.275pt}}}{RX~J1131$-$1231} & \multicolumn{2}{c}{HE~0435$-$1223} \\
Analysis & \shortstack{Omit\\dynamics} & \shortstack{Omit mass\\distribution} & \shortstack{Omit\\dynamics} & \shortstack{Omit mass\\distribution} & \shortstack{Omit\\dynamics} & \shortstack{Omit mass\\distribution} \\
\hline
\multicolumn{7}{l}{\textit{Reference, all ten metrics and ten-year light curves}} \\
All ten metrics & $-0.142^{+0.030}_{-0.004}\,(0.019)$ & $+0.022^{+0.005}_{-0.011}\,(0.023)$ & $-0.102^{+0.019}_{-0.002}\,(0.020)$ & $-0.032^{+0.016}_{-0.007}\,(0.020)$ & $-0.025^{+0.015}_{-0.005}\,(0.024)$ & $-0.053^{+0.046}_{-0.005}\,(0.025)$ \\
\hline
\multicolumn{7}{l}{\textit{Metrics excluded, ten-year light curves}} \\
Brightness SD & $-0.142^{+0.029}_{-0.002}\,(0.019)$ & $+0.023^{+0.004}_{-0.012}\,(0.023)$ & $-0.095^{+0.015}_{-0.006}\,(0.020)$ & $-0.029^{+0.013}_{-0.008}\,(0.019)$ & $-0.021^{+0.012}_{-0.007}\,(0.024)$ & $-0.066^{+0.050}_{+0.007}\,(0.024)$ \\
Detrended range & $-0.149^{+0.034}_{+0.004}\,(0.019)$ & $+0.023^{+0.005}_{-0.010}\,(0.023)$ & $-0.108^{+0.024}_{-0.001}\,(0.020)$ & $-0.034^{+0.018}_{-0.003}\,(0.020)$ & $-0.024^{+0.014}_{-0.006}\,(0.023)$ & $-0.014^{+0.011}_{-0.032}\,(0.025)$ \\
Raw range & $-0.139^{+0.027}_{-0.006}\,(0.019)$ & $+0.021^{+0.006}_{-0.010}\,(0.023)$ & $-0.095^{+0.016}_{-0.005}\,(0.020)$ & $-0.032^{+0.016}_{-0.006}\,(0.020)$ & $-0.024^{+0.012}_{-0.006}\,(0.024)$ & $-0.064^{+0.050}_{+0.006}\,(0.024)$ \\
1st-derivative tail & $-0.136^{+0.024}_{-0.003}\,(0.019)$ & $+0.024^{+0.005}_{-0.011}\,(0.024)$ & $-0.089^{+0.011}_{-0.011}\,(0.021)$ & $-0.019^{+0.011}_{-0.011}\,(0.022)$ & $-0.022^{+0.012}_{-0.011}\,(0.024)$ & $-0.024^{+0.049}_{-0.007}\,(0.027)$ \\
2nd-derivative tail & $-0.124^{+0.027}_{-0.003}\,(0.019)$ & $+0.019^{+0.007}_{-0.010}\,(0.024)$ & $-0.076^{+0.019}_{-0.008}\,(0.021)$ & $-0.034^{+0.012}_{-0.012}\,(0.020)$ & $-0.023^{+0.014}_{-0.008}\,(0.027)$ & $-0.038^{+0.043}_{-0.007}\,(0.025)$ \\
3rd-derivative tail & $-0.135^{+0.025}_{-0.007}\,(0.020)$ & $+0.011^{+0.006}_{-0.010}\,(0.024)$ & $-0.112^{+0.017}_{-0.008}\,(0.021)$ & $-0.009^{+0.008}_{-0.013}\,(0.020)$ & $-0.027^{+0.011}_{-0.007}\,(0.025)$ & $-0.073^{+0.046}_{-0.002}\,(0.025)$ \\
HME rate & $-0.140^{+0.029}_{-0.007}\,(0.020)$ & $+0.023^{+0.005}_{-0.011}\,(0.023)$ & $-0.100^{+0.018}_{-0.005}\,(0.020)$ & $-0.032^{+0.016}_{-0.006}\,(0.020)$ & $-0.025^{+0.014}_{-0.006}\,(0.024)$ & $-0.053^{+0.046}_{-0.007}\,(0.025)$ \\
Clean-HME rate & $-0.143^{+0.033}_{-0.004}\,(0.019)$ & $+0.024^{+0.005}_{-0.011}\,(0.023)$ & $-0.103^{+0.018}_{-0.003}\,(0.020)$ & $-0.030^{+0.012}_{-0.009}\,(0.019)$ & $-0.023^{+0.013}_{-0.008}\,(0.024)$ & $-0.055^{+0.048}_{-0.005}\,(0.025)$ \\
HME duration & $-0.144^{+0.034}_{-0.006}\,(0.019)$ & $+0.022^{+0.005}_{-0.011}\,(0.023)$ & $-0.101^{+0.016}_{-0.005}\,(0.019)$ & $-0.033^{+0.017}_{-0.004}\,(0.020)$ & $-0.025^{+0.014}_{-0.006}\,(0.023)$ & $-0.054^{+0.045}_{-0.005}\,(0.024)$ \\
Clean-HME duration & $-0.144^{+0.030}_{-0.001}\,(0.019)$ & $+0.021^{+0.008}_{-0.010}\,(0.022)$ & $-0.102^{+0.019}_{-0.002}\,(0.020)$ & $-0.033^{+0.017}_{-0.004}\,(0.019)$ & $-0.026^{+0.015}_{-0.005}\,(0.024)$ & $-0.053^{+0.045}_{-0.007}\,(0.024)$ \\
Duration metrics & $-0.145^{+0.035}_{-0.002}\,(0.019)$ & $+0.023^{+0.006}_{-0.011}\,(0.023)$ & $-0.102^{+0.019}_{-0.004}\,(0.019)$ & $-0.034^{+0.016}_{-0.004}\,(0.019)$ & $-0.026^{+0.015}_{-0.004}\,(0.023)$ & $-0.054^{+0.046}_{-0.007}\,(0.024)$ \\
HME metrics & $-0.147^{+0.038}_{-0.001}\,(0.020)$ & $+0.026^{+0.008}_{-0.009}\,(0.024)$ & $-0.097^{+0.017}_{-0.003}\,(0.019)$ & $-0.035^{+0.018}_{-0.006}\,(0.020)$ & $-0.026^{+0.015}_{-0.004}\,(0.024)$ & $-0.056^{+0.049}_{-0.001}\,(0.024)$ \\
Derivative metrics & $-0.010^{+0.008}_{-0.017}\,(0.028)$ & $+0.007^{+0.010}_{-0.008}\,(0.027)$ & $+0.050^{+0.001}_{-0.031}\,(0.033)$ & $+0.012^{+0.010}_{-0.011}\,(0.027)$ & $+0.001^{+0.013}_{-0.017}\,(0.038)$ & $-0.011^{+0.021}_{-0.011}\,(0.032)$ \\
All but derivatives & $-0.125^{+0.026}_{-0.007}\,(0.021)$ & $+0.030^{+0.007}_{-0.010}\,(0.023)$ & $-0.082^{+0.022}_{-0.000}\,(0.021)$ & $-0.028^{+0.014}_{-0.008}\,(0.020)$ & $-0.023^{+0.010}_{-0.008}\,(0.024)$ & $-0.038^{+0.038}_{-0.007}\,(0.025)$ \\
\hline
\multicolumn{7}{l}{\textit{Macroimage excluded, ten metrics and ten-year light curves}} \\
A & $-0.173^{+0.040}_{+0.005}\,(0.021)$ & $+0.020^{+0.008}_{-0.009}\,(0.024)$ & $-0.104^{+0.020}_{-0.006}\,(0.023)$ & $-0.020^{+0.010}_{-0.012}\,(0.022)$ & $-0.036^{+0.013}_{-0.006}\,(0.026)$ & $-0.060^{+0.033}_{-0.012}\,(0.027)$ \\
B & $-0.164^{+0.032}_{+0.003}\,(0.020)$ & $+0.023^{+0.004}_{-0.010}\,(0.022)$ & $-0.109^{+0.013}_{-0.008}\,(0.022)$ & $+0.002^{+0.005}_{-0.015}\,(0.022)$ & $-0.027^{+0.014}_{-0.008}\,(0.026)$ & $-0.022^{+0.038}_{-0.009}\,(0.027)$ \\
C & $-0.111^{+0.030}_{-0.001}\,(0.021)$ & $+0.023^{+0.006}_{-0.012}\,(0.024)$ & $-0.158^{+0.025}_{-0.005}\,(0.022)$ & $-0.035^{+0.016}_{-0.004}\,(0.022)$ & $-0.028^{+0.011}_{-0.009}\,(0.028)$ & $-0.067^{+0.032}_{-0.008}\,(0.027)$ \\
D & $-0.116^{+0.018}_{-0.016}\,(0.020)$ & $+0.024^{+0.004}_{-0.013}\,(0.024)$ & $-0.040^{+0.009}_{-0.011}\,(0.021)$ & $-0.033^{+0.016}_{-0.005}\,(0.020)$ & $-0.019^{+0.013}_{-0.008}\,(0.025)$ & $-0.023^{+0.028}_{-0.011}\,(0.027)$ \\
\hline
\multicolumn{7}{l}{\textit{Monitoring duration, all ten metrics}} \\
First five years & $-0.117^{+0.022}_{-0.008}\,(0.023)$ & $-0.005^{+0.025}_{+0.001}\,(0.023)$ & $-0.070^{+0.008}_{-0.014}\,(0.023)$ & $-0.080^{+0.038}_{-0.008}\,(0.023)$ & $-0.024^{+0.011}_{-0.013}\,(0.026)$ & $-0.093^{+0.041}_{-0.019}\,(0.027)$ \\
First seven years & $-0.132^{+0.022}_{+0.001}\,(0.023)$ & $+0.014^{+0.013}_{-0.006}\,(0.022)$ & $-0.087^{+0.012}_{-0.010}\,(0.022)$ & $-0.050^{+0.017}_{-0.010}\,(0.023)$ & $-0.026^{+0.005}_{-0.014}\,(0.025)$ & $-0.081^{+0.061}_{-0.006}\,(0.025)$ \\
\hline
\end{tabular}
\par\vspace{4pt}
\begin{minipage}{\textwidth}
\footnotesize
\textit{Note.} Entries give the baseline-corrected size offset and same-physics test RMSE (in parentheses), both in dex. Offset superscripts and subscripts give signed differences from the central estimate to the 84th and 16th bootstrap percentiles. Both can have the same sign when the original estimate lies outside the interval. The 100 whole-quad resamples repeat preprocessing, penalty selection, and all five cross-validation fits. Central estimates retain the original median fold offset and pooled test RMSE. Metric exclusions remove the corresponding mean and both scatters from all four macroimages. Macroimage exclusions refit the model using the other three macroimages. Shorter light curves use the first five or seven years of the same realizations.
\end{minipage}
\end{table}
\twocolumngrid

\bibliographystyle{aasjournalv7}
\bibliography{bib}

\end{document}